\documentclass[prl,letterpaper,aps,10pt,superscriptaddress,twocolumn,floatfix,showpacs]{revtex4-1}
\usepackage{graphicx}
\usepackage{amsmath,amssymb,amsfonts}
\usepackage{graphicx,braket,xcolor,subfigure,upgreek}
\usepackage[colorlinks, linkcolor=blue, citecolor=blue, urlcolor=blue, breaklinks=true]{hyperref}
\usepackage{microtype}
\usepackage{bbm}
\usepackage{color}
\usepackage[english]{babel}

\newcommand{\fref}[1]{Fig.~\ref{#1}}
\newcommand{\ts}[1]{_\text{#1}}

\graphicspath{{figures/}}

\begin{document}

 \title{A Nanoscale Coherent Light Source}

\author{Raphael Holzinger}
\affiliation{Institut f\"ur Theoretische Physik, Universit\"at Innsbruck, Technikerstr. 21a, A-6020 Innsbruck, Austria}
\author{David Plankensteiner}
\affiliation{Institut f\"ur Theoretische Physik, Universit\"at Innsbruck, Technikerstr. 21a, A-6020 Innsbruck, Austria}
\author{Laurin Ostermann}
\affiliation{Institut f\"ur Theoretische Physik, Universit\"at Innsbruck, Technikerstr. 21a, A-6020 Innsbruck, Austria}
\author{Helmut Ritsch}
\affiliation{Institut f\"ur Theoretische Physik, Universit\"at Innsbruck, Technikerstr. 21a, A-6020 Innsbruck, Austria}
\date{\today}

\begin{abstract}
Generically, a laser is composed of an optical resonator coupled to a gain medium. If the light amplification via stimulated emission dominates the mirror losses, the emitted light is coherent. Recent studies have shown that sub-wavelength sized rings of quantum emitters possess subradiant eigenmodes which mimic high-$Q$ optical resonators. We add a continuously pumped atom as a gain medium in the ring's center creating a minimalistic coherent light source. The system behaves like a thresholdless laser, featuring a narrow linewidth well below the natural linewidth of the constituent atoms.
\end{abstract}


\maketitle

\section{Introduction}
Conventional lasers consist of an optical cavity filled with a gain medium, typically comprised by an ensemble of energetically inverted emitters amplifying the light field via stimulated emission. Pioneering experiments have realized lasers with the most minimalistic gain medium yet, a single atom~\cite{brune1987realization,mcKeever2003experimental,davidovich1987quantum,walther1988single,an1994microlaser,astafiev2007single, rastelli2019single,loeffler1997spectral}. Corresponding theoretical quantum models have already been studied extensively for several decades~\cite{meschede1985one, mu1992one, pellizzari1994photon,salzburger2005theory}. Standard models of a single-atom laser still feature a macroscopic optical resonator supporting the corresponding laser light mode. Technically, the noise of the cavity mirrors is a substantially limiting factor for the frequency stability of a laser. This can be reduced when working in the bad cavity regime, such that the coherence is stored in the atomic dipoles rather than the light field. In such superradiant lasers~\cite{meiser2009prospects,meiser2010steady,bonnet2012,maier2014superradiant,hotter2019cooling} the properties of the emitted light are governed by the gain medium rather than the resonator.

In this work we go one step further removing the cavity altogether and consider a nano-scale system where atomic quantum emitters provide for the necessary gain while simultaneously acting as a resonator. Thus, in principle, the  size of the entire setup can be reduced to even below the order of the laser wavelength. Such a device is characterized solely by the spectral properties of the atoms.
\begin{figure}[t]
\centering
\centering\includegraphics[width=\columnwidth]{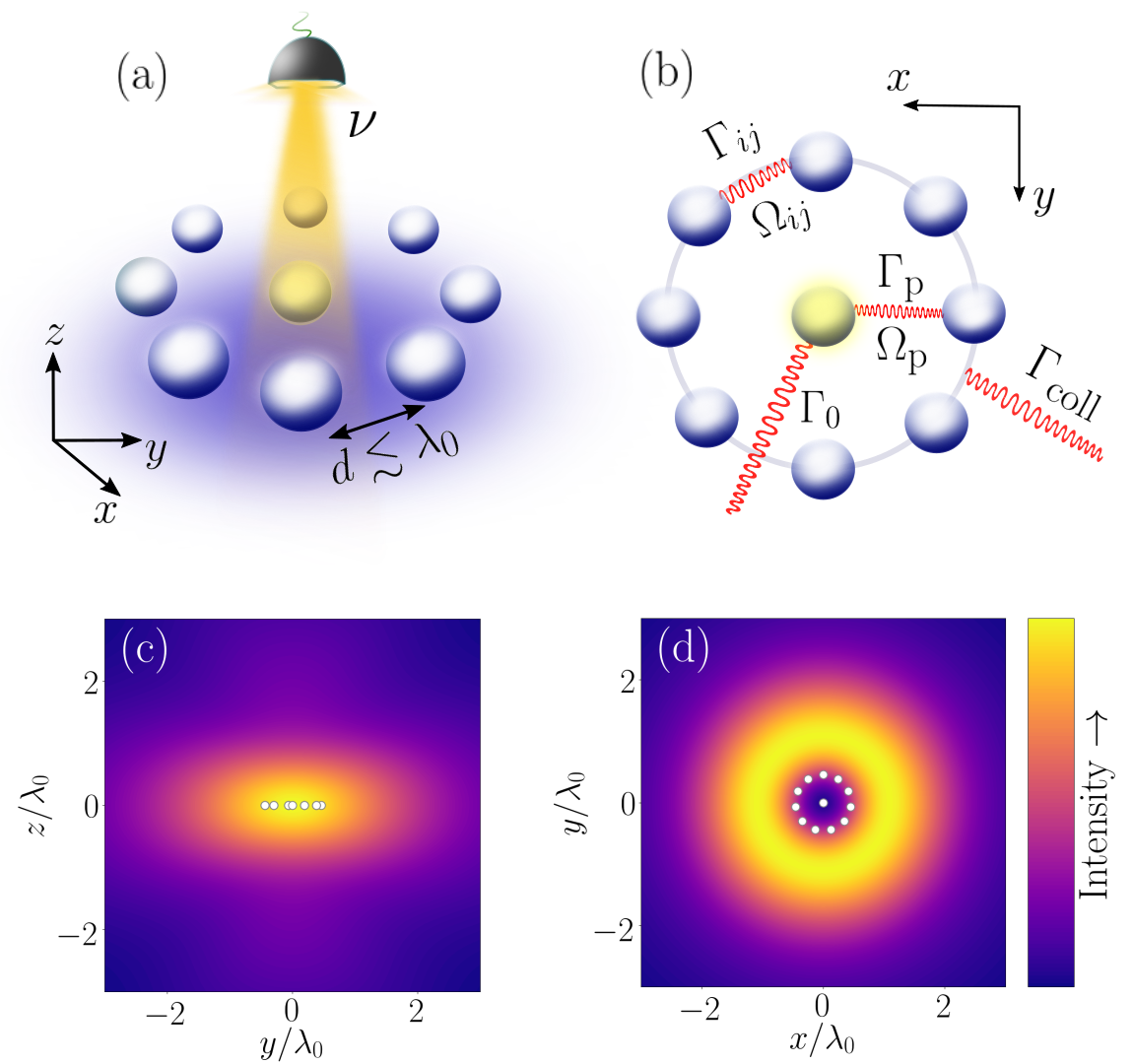}
\caption{\emph{Coherent Light Emission from a Partially Pumped Atomic Array.} (a) A ring of atoms with an additional atom in its center incoherently pumped with a rate $\nu$. (b) The atoms decay at a spontaneous decay rate $\Gamma_0$ and are collectively coupled to the center atom with dispersive coupling $\Omega\ts{p}$ and dissipative coupling $\Gamma\ts{p}$, respectively. In turn, the ring atoms have couplings $\Omega_{ij}$ and $\Gamma_{ij}$ amongst each other. The symmetric excitation exhibits a collective decay rate $\Gamma\ts{coll}$. (c) The field intensity generated in the steady state according to eq. \eqref{eq:intensity} for a ring of $N = 11$ atoms in the $xz$-plane with $y = 2.5\lambda_0$ and interatomic distance $d=\lambda_0/5$ and pumping rate $\nu = 0.1\Gamma_0$. (d) The field intensity in the $xy$-plane with $z = 2.5\lambda_0$.}
\label{model}
\end{figure}

As discovered recently, tailored dipole-coupled atomic arrays possess collective eigenmodes with a very long lifetime demonstrating analogous characteristics to a high-$Q$ optical cavity mode~\cite{moreno2019extraordinary,manzoni2018optimization}. Such arrangements could be implemented, e.g.\ by means of optical tweezers~\cite{Barredo2016atom,Barredo2018synthetic,wang2019preparation} or superconducting qubit setups operating in the microwave regime~\cite{blais2004cavity}. We study the prospects of implementing a minimalistic sub-wavelength sized laser by incoherently pumping some of the dipoles in such a nano array. As our generic setup we consider a single atom placed in the center of a small ring comprised of identical emitters. The collective coupling to the other emitters in the ring is mediated by virtual photon exchange through the electromagnetic vacuum~\cite{lehmberg1970radiation,hood2016atom,astafiev2007single}. The collective eigenmodes of the outer ring take on the role of a resonator mode.

We show that such a minimal model constitutes a steady-state coherent light source with a spectral linewidth well below the single atom decay rate. Therefore, it can be viewed as a minimal implementation of a laser. Depending on the number of atoms and the configuration of the array, the collective nature of the dipole-dipole couplings leads to strong quantum correlations within the atoms and an inherent emission of a coherent field. Optimal operation is achieved when the collective state in the ring atoms features a single subradiant excitation only.

\section{Model}
We consider $N$ identical two-level atoms with excited state $\ket{e}$ and ground state $\ket{g}$ each, separated in frequency by $\omega_0$ and arranged in a ring geometry at an inter-atomic distance of $d \lesssim \lambda_0 = 2\pi c/\omega_0$. An additional gain atom is placed in the center of the ring as depicted in~\fref{model}a and is assumed to be pumped to its upper level incoherently at a rate $\nu$ (after having eliminated auxiliary levels). The corresponding raising (lowering) operators of the $i$th atom are $\sigma^\pm_i$ for $i \in \lbrace 1, 2, \ldots, N, p \rbrace$ (the index $p$ corresponds to the central, pumped atom). The excited state is subject to spontaneous emission with a rate $\Gamma_0$. All transition dipoles $\pmb{\mu}_i$ are chosen such that they point in $z$-direction.

At the considered distances, the fields emitted by each of the atoms interfere resulting in effective dipole-dipole interactions~\cite{lehmberg1970radiation}, so that the atomic ring acts like a resonator~\cite{moreno2019extraordinary} coupled to the gain atom in its center.

Using standard quantum optical techniques~\cite{gardiner2004quantum} we obtain a master equation for the internal dynamics of the emitters,
$ \dot{\rho} = i \left[ \rho, H \right] +\mathcal{L}_\Gamma \left[\rho \right] + \mathcal{L}_\nu \left[\rho \right],
$
where the Lindblad term describing the incoherent pumping of the central atom is given by
$
\mathcal{L}_\nu \left[ \rho \right] =  \frac{\nu}{2}\left(2 \sigma^+_\mathrm{p} \rho \sigma^-_\mathrm{p} - \sigma^-_\mathrm{p} \sigma^+_\mathrm{p} \rho -\rho \sigma^-_\mathrm{p} \sigma^+_\mathrm{p} \right).
$
The corresponding Hamiltonian in a frame rotating at the atomic transition frequency $\omega_0$ is
\begin{equation} \label{eq:hamiltonian} 
  H = \sum_{i,j:i \neq j} \Omega_{ij} \sigma^+_i \sigma^-_j,
\end{equation}
while the Lindblad operator accounting for collective spontaneous emission reads
\begin{equation} \label{eq:lindblad_gamma} 
\mathcal{L}_\Gamma \left[ \rho\right] = \sum_{i,j} \frac{\Gamma_{ij}}{2}\left(2 \sigma^-_i\rho \sigma^+_j -\sigma^+_i \sigma^-_j \rho - \rho \sigma^+_i \sigma^-_j \right).
\end{equation}

The collective coupling rates $\Omega_{ij}$ and $\Gamma_{ij}$ are given as the real and imaginary part of the overlap of the transition dipole of the $i$th atom with the electric field emitted by the $j$th atom (see Eq.~\eqref{eq:greens_tensor}).

The emitted electric field $\pmb{E}^+(\pmb{r})$ can be used to compute the field intensity as (see Eq.~\eqref{eq:field}), i.e.\
\begin{equation} \label{eq:intensity} 
I(\pmb{r}) = \left \langle \pmb{E}^+(\pmb{r}) \pmb{E}^-(\pmb{r}) \right \rangle,
\end{equation}

The steady-state intensity is shown in~\fref{model}c and~\fref{model}d for typical operating conditions.

\section{Continuous Collective Emission}
Our goal is to find operating regimes where the system emits coherent light with a narrow linewidth. As the configuration is symmetric with respect to the coupling of the ring atoms to the gain atom in the center, we can expect the ring atoms to be driven into a symmetric excitation state given as
\begin{equation}
\ket{\psi\ts{sym}}  = \frac{1}{\sqrt{N}}\sum_{j=1}^N \sigma^+_j \ket{g}^{\otimes N}.
\end{equation}

\begin{figure}[ht]
\centering
    \includegraphics[width=\columnwidth]{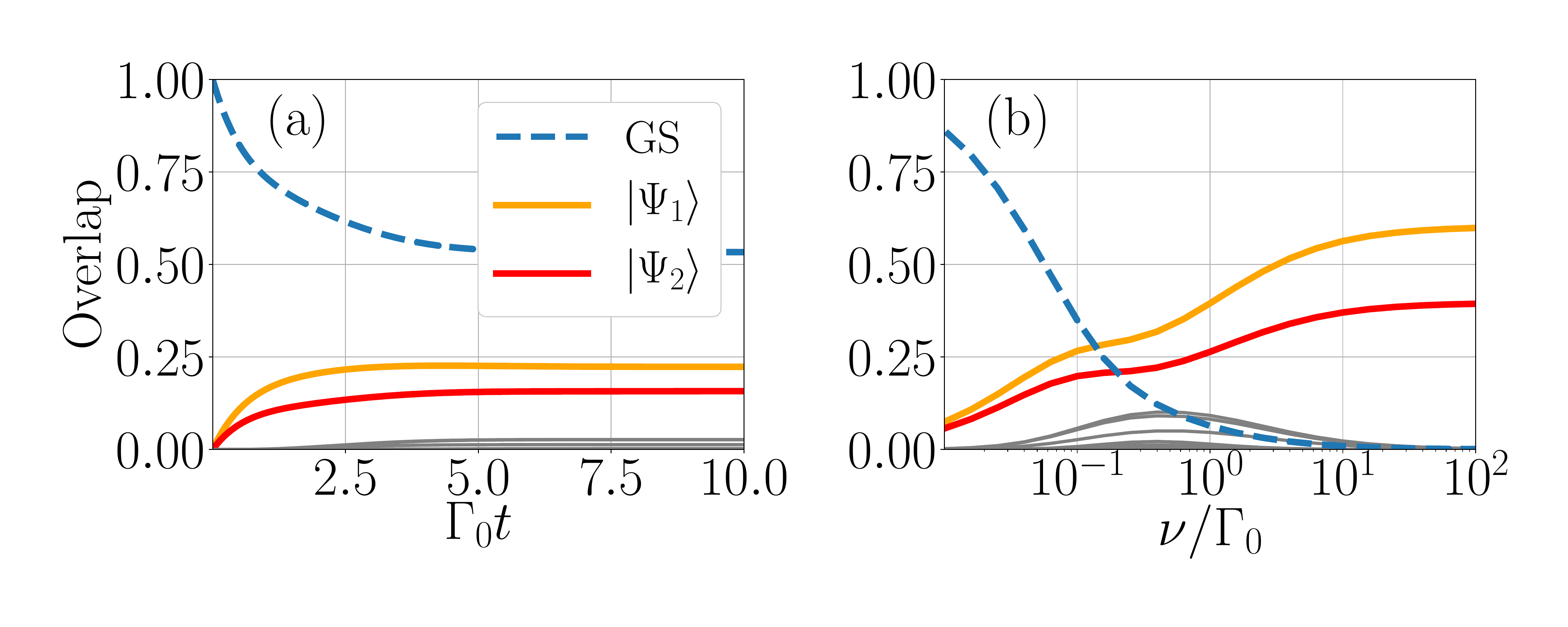}
    \caption{\emph{Dissipative System Dynamics.} (a) Time evolution of the population of the eigenstates of the Hamiltoniann from eq.~\eqref{eq:hamiltonian}]for $N=5$ ring atoms with an interatomic distance $d=\lambda_0/2$ and an incoherent pump rate $\nu=\Gamma_0/2$ starting from the ground state. The state $\ket{\Psi_{1,2}}$ feature a large contribution from the symmetric state of the ring atoms $\ket{\psi\ts{sym}}$ and show significantly higher populations than all other excited eigenstates (gray lines) at all times. (b) Stationary population of the eigenstates for different pump rates. We can see that even for large pump rates $\nu>\Gamma_0$ the symmetric single-excitation states dominate.}
    \label{symmetric_state}
\end{figure}

In accordance with standard laser theory we will target parameters for which a symmetric excitation of the ring atoms constitutes a good cavity, i.e.\ the radiative loss is sufficiently small. To this end, we study the stationary populations of different eigenstates of our Hamiltonian from Eq.~\eqref{eq:hamiltonian} during a time evolution starting from the ground state as depicted in~\fref{symmetric_state}a.

Indeed, as shown in~\fref{symmetric_state}, we find that the two eigenstates involving the symmetric single-excitation state in the ring are occupied predominately at all times (except for the ground state). These states are given by
\begin{align}\label{eq:dominant_states} 
\ket{\Psi_{i}} = a_i \ket{g}^{\otimes N }\otimes \ket{e} + b_i \ket{\psi\ts{sym}}\otimes\ket{g},
\end{align}
for $i \in \lbrace 1, 2 \rbrace$, where $a_i$ and $b_i$ depend on the particular geometry with $\left| a_i \right|^2 + \left| b_i \right|^2 = 1$.

Note that the gain atom can only emit one photon into the ring at a time. Hence, the single-excitation manifold dominates the dynamics even for pump rates substantially larger than the single-atom decay rate. This is shown in~~\fref{symmetric_state}b, where we plot the occupation probability of different eigenstates at steady state as a function of $\nu$.

The fact that the ring does indeed form a resonator can be seen more clearly as follows. Let us assume that only the symmetric state in the ring is populated. Thus, we can rewrite the Hamiltonian in the subspace spanned by the ground and excited state of the gain atom in the center, as well as the ground state of the ring and its symmetric state obtaining (see Appendix)
\begin{equation}\label{eq:Hamiltonian_sym}
H\ts{sym} = \Omega\ts{sym} \sigma\ts{sym}^+ \sigma\ts{sym}^-  + \sqrt{N}\Omega_\mathrm{p} \left(\sigma\ts{sym}^+\sigma_\mathrm{p}^{-} + \text{H.c.}\right),
\end{equation}
where $\Omega\ts{sym}=\sum_{j=2}^N \Omega_{1j}$ is the dipole energy shift of the symmetric state. Written like this, the Hamiltonian resembles the Jaynes-Cummings Hamiltonian with the ring taking on the role of the cavity mode. In this sense, the symmetric subspace lowering operator
$
\sigma\ts{sym}^- := \ket{g}^{\otimes{N}}\bra{\psi\ts{sym}}\otimes\mathbbm{1}\ts{p}
$
can be interpreted as the photon annihilation operator of our "cavity". The coupling between the gain atom and the cavity is then determined by $\Omega_\mathrm{p}$.

If we neglect the dissipative coupling between the central atom and the atoms forming the ring, i.e.\ $\Gamma_\mathrm{p} = 0$, we can rewrite the decay of the system as
$
\mathcal{L} \left[ \rho \right] = \mathcal{L}_\nu \left[ \rho \right] + \mathcal{L}_0 \left[ \rho \right] + \mathcal{L}\ts{sym} \left[ \rho \right],
$
with
\begin{subequations}
\begin{align}
\mathcal{L}_0 \left[ \rho \right] &= \frac{\Gamma_0}{2}\left(2\sigma_\mathrm{p}^{-}\rho\sigma_\mathrm{p}^{+} - \sigma_\mathrm{p}^{+}\sigma_\mathrm{p}^-\rho - \rho\sigma_\mathrm{p}^{+}\sigma_\mathrm{p}^-\right),
\\
\mathcal{L}\ts{sym}\left[ \rho \right] &= \frac{\Gamma\ts{sym}}{2} \left(2\sigma\ts{sym}^-\rho\sigma\ts{sym}^+ - \sigma\ts{sym}^+\sigma\ts{sym}^-\rho \right.\\
&- \left.\rho\sigma\ts{sym}^+\sigma\ts{sym}^-\right).
\notag
\end{align}
\end{subequations}
\begin{figure}[t]
  \centering
    \includegraphics[width=1\columnwidth]{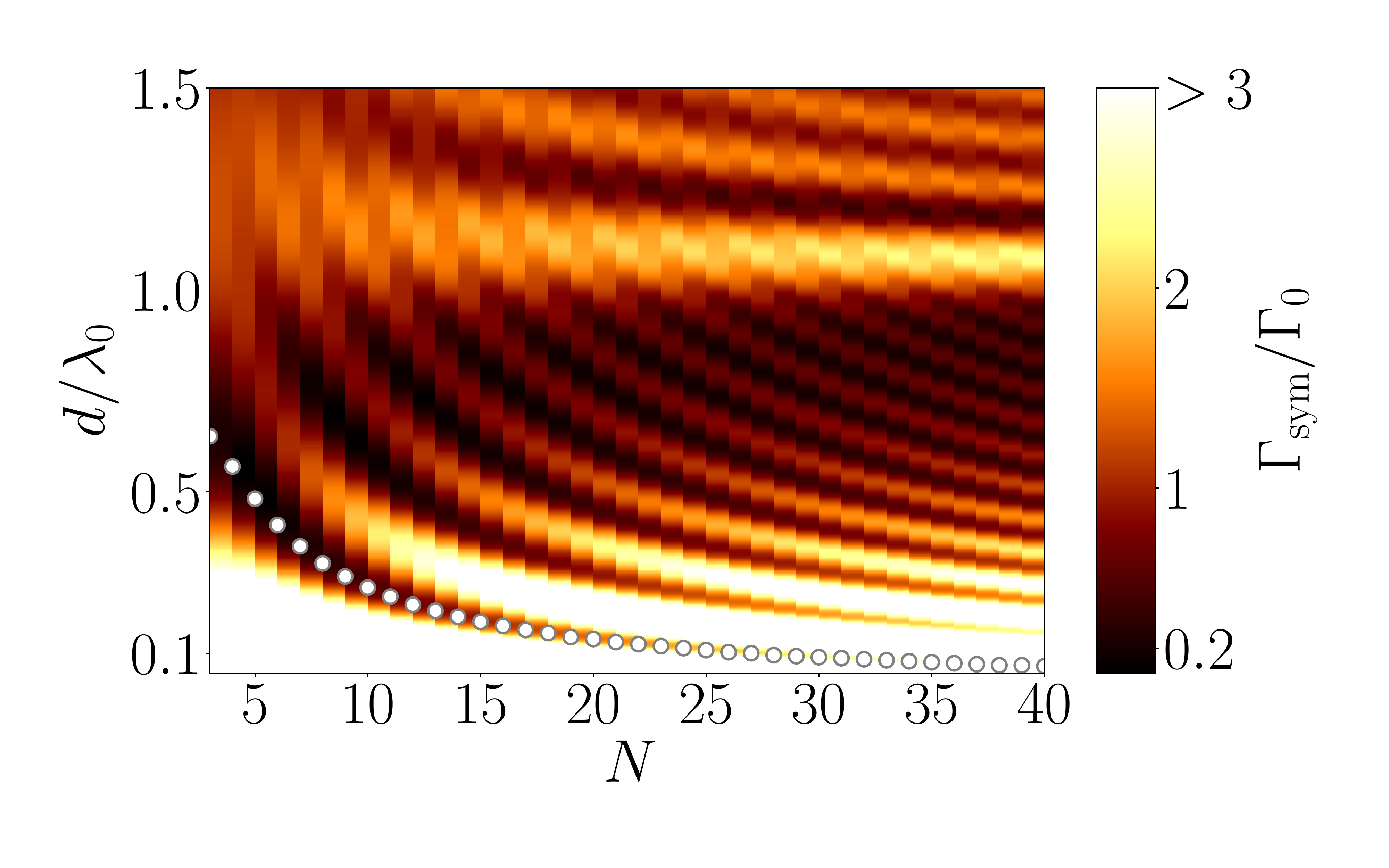}
      \vspace*{-0.8cm}
    \caption{\emph{Super- and Subradiance of the Symmetric State}. The decay rate of the symmetric state $\Gamma\ts{sym}$ as a function of the atom number in the ring and their interatomic distance. The white dots highlight specific interatomic distances where the decay of the symmetric state is the smallest (subradiant).}
    \label{symmetric_decay}
\end{figure}

\begin{figure}[ht]
  \centering
  \hspace*{-0.6cm}
    \includegraphics[width=\columnwidth]{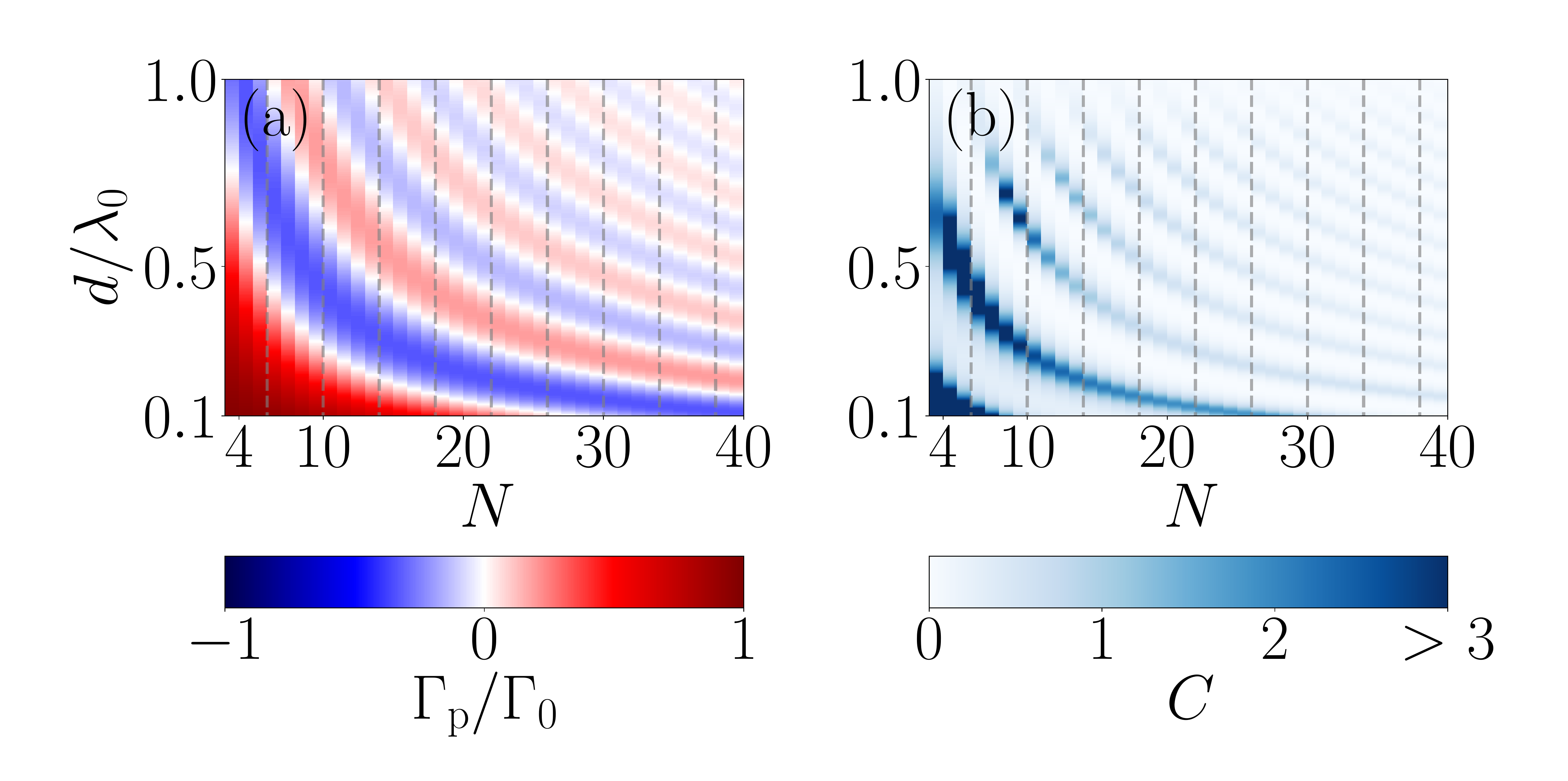}
    \vspace*{-0.4cm}
    \caption{\emph{Coupling of the Central Gain Atom to the Outer Ring.} (a) The dissipative coupling $\Gamma_\mathrm{p}$ between the central atom and the ring atoms is plotted as a function of the atom number $N$ and the inter-atomic distance $d$. One can see that it becomes negligible at the points where $\ket{\psi\ts{sym}}$ is subradiant. (b) Cooperativity $C$ for different distances and atom numbers. The cooperativity is large when $d\to 0$ due to the divergent behavior of $\Omega_\mathrm{p}$, or when $\Gamma\ts{sym}$ is small.}
    \label{decay_rates}
\end{figure}

Minimizing the decay rate of the ring atoms is important, but in order to build up population within the ring we need an efficient coupling to the gain atom as well. In analogy to the Jaynes-Cummings model we thus define a cooperativity parameter (see Appendix)
$
C := N\Omega_\mathrm{p}^2/\left(\Gamma_0\Gamma\ts{sym}\right).
$
An efficient coherent coupling of the ring atoms to the gain atom is achieved when $C>1$. As we can see in~\fref{decay_rates}b, we reach this limit at extremely small distances or at a distance where $\Gamma\ts{sym}$ is minimal (see~\fref{symmetric_decay}). The cooperativity becomes large at $d<0.1\lambda_0$ since for $d \to 0$ the coherent coupling diverges. Yet, this is also the limit where the energy difference $\Omega\ts{sym}$ is large, which detunes the ring atoms from the gain atom. Furthermore, as we will show later, due to the superradiant loss of the ring in this limit the emitted light  features thermal statistics rather than coherence. Consequently, we find that the optimal parameter regime indeed lies where the ring atoms show a subradiant behaviour, i.e.\ at the points highlighted in~\fref{symmetric_decay}.

As seen in~\fref{decay_rates}a, the dissipative coupling of the central atom vanishes at points where the symmetric state shows suppressed spontaneous emission (see~\fref{symmetric_decay}). Hence, the loss during the excitation transport  from the gain medium to the ring is reduced as well.

\section{Photon Statistics and Spectral Properties}
We have now identified a regime where our system resembles the typical setup of a single-atom laser. In order to study the statistical properties of the emitted light we calculate the normalized second-order correlation at zero time delay $g^{(2)}(0)$ of the electric field intensity. In the far-field $r \gg \lambda_0$, where the intensity correlation function becomes independent of the position (see Appendix) and is given by
\begin{equation}\label{eq:13}
g^{(2)}(0) = \frac{\sum_{ijkl} \left \langle \sigma^+_i \sigma^+_j \sigma^-_k \sigma^-_l \right \rangle_{}}{|\sum_{mn} \left \langle \sigma^+_m \sigma^-_n \right \rangle|^2}.
\end{equation}

Coherent light exhibits a Poissonian statistic implying $g^{(2)}(0)=1$~\cite{gardiner2004quantum,mandel_wolf_1995}. Therefore, an operation in the previously identified parameter regimes leads to the emission of coherent light. In addition , we calculate the amount of emitted light, i.e.\
$
I\ts{out} := \sum_{ij} \Gamma_{ij} \left \langle \sigma^{+}_i \sigma^{-}_j \right \rangle.
$

\begin{figure}[t]
  \centering
    \includegraphics[width=\columnwidth]{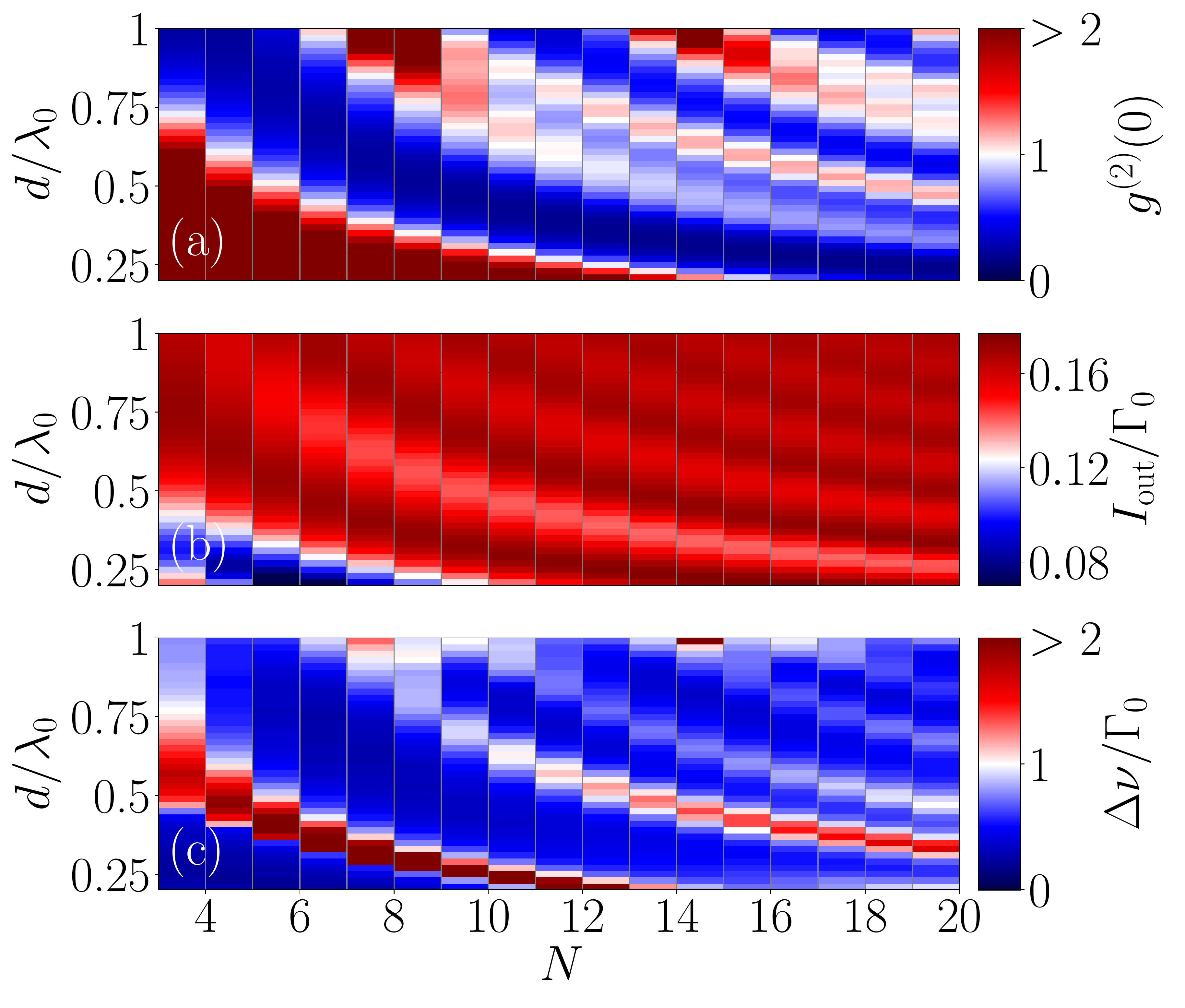}
    \caption{\emph{Intensity and Statistics of the Emitted Light.} (a) Steady-state second-order correlation as a function of the ring atom number and atom spacing. For each atom number $N$ there are specific interatomic distances $d$ where the emitted light changes from thermal-like light emission (red), passing over regions of Poissonian statistics (white), to sub-Poissonian properties (blue). (b) The radiated intensity $I\ts{out}$ for the same parameter region. Where $g^{(2)}(0) = 1$ the intensity is maximal, regardless of the atom number. (c) The spectral linewidth $\Delta \nu$ for the same parameters. It reduces to well below $\Gamma_0$. The pump rate was $\nu=0.1\Gamma_0$.}
    \label{g2_plot}
\end{figure}

In~\fref{g2_plot}a we can see that points of coherent light emission where $g^{(2)}(0)=1$ are achieved along a curve strongly resembling the optimal subradiance parameters shown in~\fref{symmetric_decay}. The points where $g^{(2)}(0)=0$ correspond to the situation where the gain atom decouples from the cavity atoms, since then only the single atom in the center can emit light. And, because it is not possible for a single atom to emit more than one photon at a time, we observe anti-bunching. However, this regime does not coincide with "lasing" since the ring atoms are not occupied. Simultaneously, the intensity shown in \fref{g2_plot}b is small, but still finite when the emitted light is coherent. This is because coherences can only build up when the loss from the atoms in the ring is sufficiently low ($\Gamma\ts{sym}$ is small), which also reduces the amount of light emitted.

In order to analyze the emitted light in more detail, we compute its spectral linewidth. Therefore, we calculate the emission spectrum by means of the Wiener-Khinchin theorem (see eq. \eqref{eq:spectrum})~\cite{carmichael2009open}. It is given as the Fourier transform of the first-order coherence function,
$
g^{(1)}(\tau):=\sum_{i,j}\left \langle \sigma_i^+(\tau) \sigma^-_j \right \rangle.
$
The spectrum has a Lorentzian shape, thus we compute the linewidth $\Delta\nu$ as the full width at half maximum (FWHM). In~\fref{g2_plot}c, we show the linewidth as a function of $N$ and the interatomic distance $d$. Once again, we find that the linewidth is small ($\Delta \nu < \Gamma_0$) at the points where the symmetric state is subradiant. It can be seen that in order to maintain coherent light emission the interatomic distances need to become smaller for an increasing number of atoms in a ring of constant radius.

Note, that in order to treat larger atom numbers in the above calculations we have truncated the Hilbert space at the second-excitation manifold (see Appendix). Since the single-excitation subspace usually dominates [as shown in \fref{symmetric_state}], neglecting any state containing more than two excitations is well justified.

\section{Threshold-Less Behavior}
In standard lasing models, coherent output light is achieved  from a certain input power threshold on. Above threshold, the intensity of the emitted light increases drastically. In an effort to identify such a threshold in our setup, we compute the properties of the output light as a function of the pump strength of the gain atom.
\begin{figure}[ht]
  \centering
    \includegraphics[width=1.0\columnwidth]{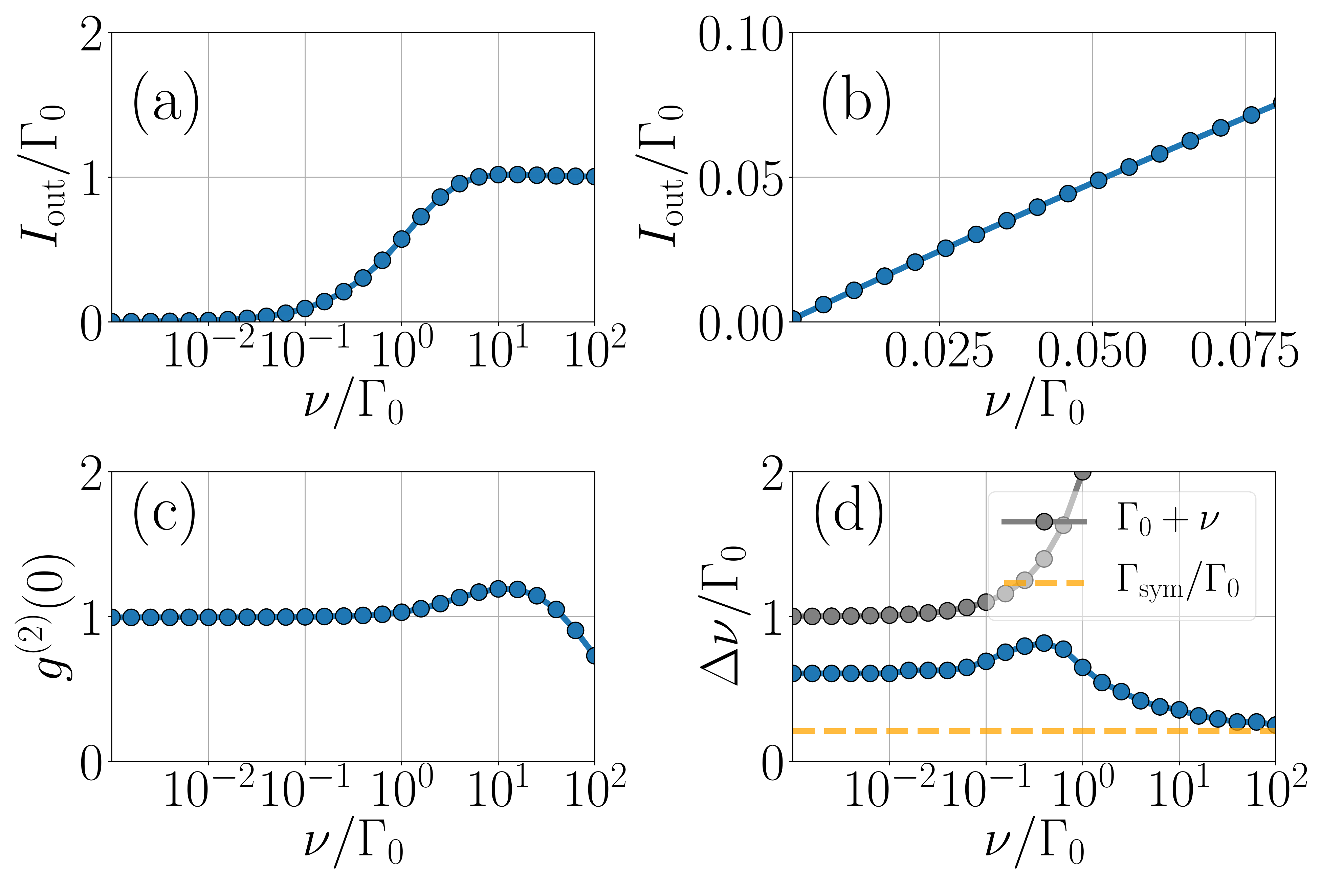}
    \caption{\textit{Threshold-Less Coherent Light Emission.} (a) $I\ts{out}$ as a function of the pump rate $\nu$ for $N=5$, $d=\lambda_0/2$ exhibiting a maximum from $\nu \approx 4\Gamma_0$ onwards. (b) A  zoom in to the weak pump region shows the immediate onset of the intensity $I_\mathrm{out}$ at small $\nu$. (c) The second-order correlation $g^{(2)}(0)$ in steady state is $1$ for finite, but small $\nu$. (d) The radiative linewidth $\Delta \nu$ (blue) in the steady state stays well below the pump broadened linewidth $\Gamma_0 + \nu$ of a single emitter (gray), and approaches the decay rate $\Gamma_\mathrm{sym}$ of the symmetric state (yellow, dashed line).}
     \label{output_field}
\end{figure}

The system does not exhibit a threshold. Such a threshold-less behavior has been observed in single-atom lasing setups~\cite{mcKeever2003experimental}. As we can see in Figs.~\ref{output_field}a and~\ref{output_field}b, the output intensity  grows as soon as the pump rate becomes nonzero, rather than requiring a sufficiently large pump rate. At the same time, the photon statistics of the emitted field are Poissonian, i.e. $g^{(2)}(0)=1$, for arbitrarily low pumping rates (see~\fref{output_field}c). The only point at which the photon statistics change is when the pump rate becomes large, $\nu\sim 10\Gamma_0$, such that the emitted light starts to reproduce the thermal statistics of the input field. It can also be seen in~\fref{output_field}a that above this point the output intensity is actually reduced. As one would expect, the linewidth of the emitted field is small ($\Delta \nu < \Gamma_0$) as long as the light is coherent (see~\fref{output_field}d). When the incoherent pumping rate $\nu$ is increased, states outside the symmetric subspace are occupied, which leads to a slight increase in the linewidth. However, by increasing $\nu$ further, the linewidth decreases again and approaches $\Gamma_\mathrm{sym}$ as the central atom decouples from the ring atoms the light is emitted from the ring in the subradiant symmetric state.

\section{Conclusions}
We predict that a continuously pumped single atom surrounded by a nano-ring of identical atoms could act as a minimal, sub-wavelength sized implementation of a laser. Under suitable operating conditions the system will emit spatially and temporarily coherent light with Poisson statistics. Our analysis reveals a close analogy to the Jaynes-Cummings model, where the outer ring atoms take on the role of a high-$Q$ cavity mode with the central atom providing for gain. The system works best when driven into a collective subradiant state with a single excitation. In this limit, spontaneous emission is suppressed and the operation strongly resembles the behavior of a threshold-less laser~\cite{Boca04}. While the implementation of such a system in a pure form could be envisioned in optical tweezer arrays of neutral atoms~\cite{Barredo2016atom}, analogous setups based on quantum dots have been implemented and are already operational in the pulsed excitation regime~\cite{le2018colloidal}.

Let us note here that there are no principal lower physical limits on the size of the system apart from the technical implementation of the structure and its pumping. Hence, very high density arrays of such lasers on a surface are possible.

\begin{acknowledgements}
We acknowledge funding from the European Union's Horizon 2020 research and innovation program under Grant Agreement No. 820404 iqClock (R.~H., D.~P. and H.~R.) as well as from the Austrian Science Fund under project P29318-N27 (L.~O.). The numerical simulations were performed with the open-source framework QuantumOptics.jl~\cite{kramer2018quantumoptics} and the graphs were produced with the open-source library Matplotlib~\cite{hunter2007matplotlib}.
\end{acknowledgements}

\bibliography{AtomLaserRef}

\section{Appendix}
\subsection{Green's Tensor}
The collective coupling rates $\Omega_{ij}$ and $\Gamma_{ij}$ are given as the real and imaginary part of the overlap of the transition dipole of the $i$th atom with the electric field emitted by the $j$th atom, i.e.\
\begin{subequations}
\begin{align} \label{eq:couplings1} 
\Omega_{ij} &= -\frac{3\pi \Gamma_0}{k_0}\Re \left( {\pmb{\mu}}_i^*\cdot \pmb{G} \left( \pmb{r}_i-\pmb{r}_j,\omega_0 \right)\cdot \pmb{\mu}_j \right),
\\
\label{eq:couplings2} 
\Gamma_{ij} &= \frac{6\pi \Gamma_0}{k_0} \Im \left( \pmb{\mu}_i^*\cdot \pmb{G} \left( \pmb{r}_i-\pmb{r}_j,\omega_0 \right)\cdot \pmb{\mu}_j \right).
\end{align}
\end{subequations}

In the above, $\pmb{G} \left(\pmb{r},\omega_0 \right)$ is the electromagnetic Green's tensor of an oscillating dipole source in the vacuum~\cite{jackson2007classical} which reads
\begin{equation} \begin{aligned}\label{eq:greens_tensor} 
\pmb{G} \left(\pmb{r},\omega_0 \right)\cdot \pmb{\mu} &= \frac{e^{i k_0 r}}{4\pi r}\Big[ \left(\hat{\pmb{r}} \times \pmb{\mu} \right) \times \hat{\pmb{r}} +
\\
&+ \left(\frac{1}{k_0^2 r^2} - \frac{i}{k_0 r}\right) \left(3\hat{\pmb{r}} \left( \hat{\pmb{r}}\cdot \pmb{\mu} \right)- \pmb{\mu} \right) \Big],
\end{aligned} \end{equation}
where $r = \left| \pmb{r} \right|$ and $\hat{\pmb{r}} = \pmb{r}/r$ is the position unit vector, $k_0 = \omega_0/c$ and $\Gamma_0 = \left|\pmb{\mu}\right|^2 k_0^3/ \left( 3\pi \epsilon_0 \right)$ is the spontaneous emission rate of a single atom.

The electric field operator can be obtained directly from the atomic operators~\cite{PhysRevX.7.031024,PhysRevA.95.033818} as
\begin{equation} \label{eq:field} 
\pmb{E}^+(\pmb{r}) = \frac{|\pmb{\mu}|k_0^2}{\epsilon_0}\sum_i \pmb{G}\left((\pmb{r}-\pmb{r}_i, \omega_0 \right)\cdot \pmb{\mu}_i \, \sigma^-_i
\end{equation}
for the positive frequency component. This is the electric field generated by an ensemble of $N$ atoms at the position $\pmb{r}$ in the vacuum.

In order to arrive at the expression for the normalized second-order correlation function $g^{(2)}(0)$ in the far-field we use the fact, that the overlap of the Green's tensor with the atomic dipole becomes approximately independent for $\left| \pmb{r}-\pmb{r}_i \right|\gg \lambda_0$. In our case of identical two-level emitters polarized in $z$-direction distributed in the $xy$-plane qq.~\eqref{eq:greens_tensor} simplifies to
\begin{equation} \label{eq:green_simple} 
\pmb{G} \left(\pmb{r}-\pmb{r}_i,\omega_0 \right)\cdot \pmb{\mu}_i \approx \frac{e^{ikr}}{4\pi r} \hat{e}_z \Big(1-\frac{1}{k_0^2 r^2}-\frac{i}{k_0 r}\Big),
\end{equation}
where $\left| \pmb{r}-\pmb{r}_i \right| \gg \lambda_0$ and therefore upon normalization the second-order correlation function becomes independent of $\pmb{r}$ and $\pmb{r}_i$.

\subsection{Symmetric Subspace}
As mentioned in the main text, during the whole time evolution and for any incoherent pumping rate $\nu$ the ring is mainly in the symmetric state. This allows us to restrict ourselves to a subspace within the single-excitation manifold where either the central atom is excited or the symmetric state of the ring is populated. The Hilbert space is spanned by these two states and the ground state of the system, i.e.\
\begin{equation} \label{subspace}
\Big\{ |\phi_1 \rangle,|\phi_2 \rangle,|\phi_3 \rangle \Big\}\equiv \Big\{ |\psi_\mathrm{sym}\rangle \otimes |g\rangle,|g\rangle^{\otimes N}\otimes |e\rangle,|g\rangle^{\otimes N}\otimes |g\rangle\Big\}.
\end{equation}

Within this subspace the nonzero matrix elements of the Hamiltonian are given by
\begin{subequations}\label{eq:expect_ham}
\begin{align}
\langle \phi_1| H |\phi_1\rangle &= \Omega_{\mathrm{sym}},
\\
\langle \phi_1| H |\phi_2 \rangle &= \sqrt{N}\Omega_{\mathrm{p}}.
\end{align}
\end{subequations}

In turn, this allows us to rewrite the Hamiltonian in this basis as
\begin{align}\label{eq:Hamiltonian_sym_appendix}
H\ts{sym} &= \Omega\ts{sym}\sigma\ts{sym}^+\sigma\ts{sym}^- + \sqrt{N}\Omega_\mathrm{p}\left(\sigma\ts{sym}^+\sigma_\mathrm{p}^{ge} + \text{H.c.}\right),
\end{align}
where the subspace lowering operator is given by
\begin{align}\label{new_transitions}
\sigma\ts{sym}^- = \ket{g}^{\otimes{N}}\bra{\psi\ts{sym}}\otimes \mathbbm{1}_p.
\end{align}

If there is only a single excitation present in the system the Lindblad operator accounting for the collective spontaneous emission can be rewritten as
\begin{align}
\mathcal{L}_\Gamma[\rho] = \sum_{i,j} \frac{\Gamma_{ij}}{2}\Big[\sigma^{eg}_i \sigma^{ge}_j,\rho \Big],
\end{align}
 where the atomic density matrix in the steady state will live in the symmetric subspace, such that
 \begin{align}
 \rho \propto \sum_{i,j}^2 \left| \phi_i \right \rangle \left| \langle \phi_j \right|.
 \end{align}

Applying the Lindblad superoperator will yield the decay rates $\Gamma_\mathrm{p}$, $\Gamma_\mathrm{sym}$ and $\Gamma_\mathrm{0}$ for $i \neq j$, $i,j=1$ and $i,j=2$, respectively. The collective decay $\Gamma_\mathrm{p}$ between the central atom and the ring atoms will be approximately zero for the distances where $g^{(2)}(0) = 1$ and can be neglected as is discussed in the main text. The ring features the collective decay $\Gamma_\mathrm{sym} = \langle \psi_\mathrm{sym}|\mathcal{L}_\Gamma[\rho]|\psi_\mathrm{sym}\rangle$ and the central atom independent spontaneous emission $\Gamma_0$ with the decay operators $\sigma^-_\mathrm{sym}$ and $\sigma^-_\mathrm{p}$ respectively. Therefore the Lindblad term can be split into
\begin{align}\label{eq:lindblad_single_ex}
\mathcal{L}_\Gamma[\rho] = \mathcal{L}_{\Gamma_\mathrm{sym}}[\rho]+\mathcal{L}_{\Gamma_0}[\rho].
\end{align}

This leads to a form of the Hamiltonian and Master equation which resembles the Jaynes-Cummings model, where the good cavity is given by the subradiant symmetric state of the ring atoms.

The definition of the cooperativity parameter can be understood as follows. As can be seen in $H_\mathrm{sym}$, the coupling coefficient between the central atom and one of the ring atoms is given by $\sqrt{N} \Omega_\mathrm{p}$, whereas the spontaneous decay into the vaccum modes for the center atom is simply $\Gamma_0$. As analyzed in the main text, the ring atoms are predominately in the symmetric state with a collective decay rate $\Gamma_\mathrm{sym}$ for the parameters where the symmetric state is maximally subradiant. Interpreting the ring in the symmetric state as a cavity and the center atom as the gain medium leads to the definition of the cooperativity parameter $C$.

\begin{figure}[ht]
  \centering
    \includegraphics[width=\columnwidth]{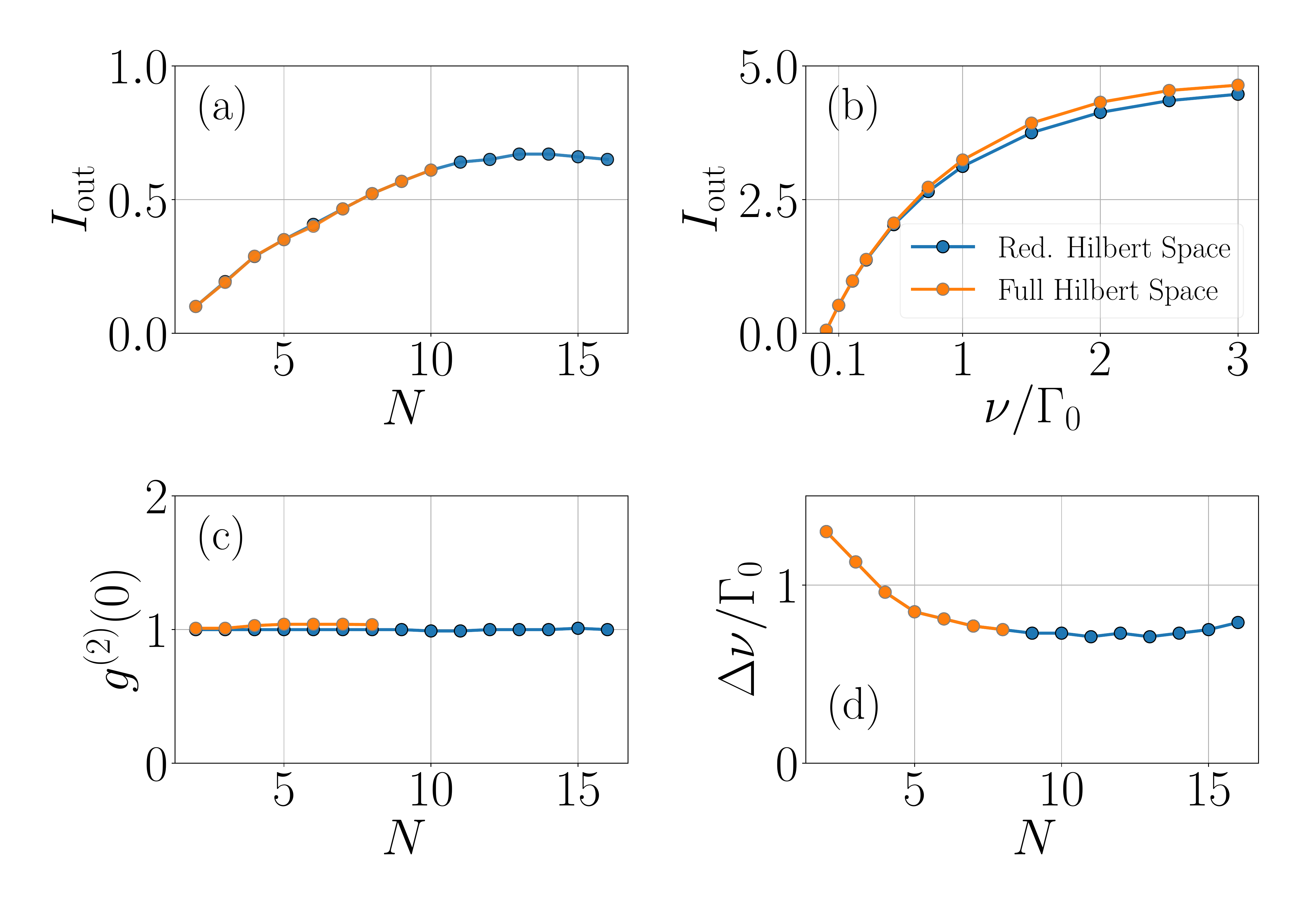}
    \caption{(a) Scaling behaviour of $I_\mathrm{out}$ in the steady state as a function of the atom number in the ring, where the interatomic distance for each $N$ is chosen along the white circles in fig.\ 3 of the main text and the incoherent pumping rate $\nu = 10^{-1}\Gamma_0$. (b) Comparison of the Master equation with the cut-off at the second excitation manifold as a function of the incoherent pumping rate $\nu$ for $N=8$ atoms in the ring. (c,d) The intensity correlation function $g^{(2)}(0)$ and the linewidth $\Delta \nu$ as a function of $N$ for a pumping rate $\nu = 10^{-1}\Gamma_0$ and an interatomic distance $\lambda_0/2$ between neighbouring atoms.}
    \label{scaling}
\end{figure}

\subsection{Truncating the Hilbert Space at Low Excitation}
Concerning the scaling behaviour of the system a mean field treatment even with the inclusion of correlations to second order is not sufficient since the properties of the steady state strongly depend on correlations of higher orders in particular $g^{(2)}(0)$ involves products of four operators. In order to analyze the scaling behaviour for a larger number of emitters in the ring we restrict the system to two excitations. This cut-off can only be a good approximation to the full model for small enough incoherent pumping rates $\nu$. In fig.~\ref{scaling}b the output intensity $I_\mathrm{out}$ in the steady state of the reduced Hilbert space is compared to the full model for eight atoms in the ring and a good agreement can be found for $\nu \leq \Gamma_0$. For $I_\mathrm{out}$, the intensity correlation function $g^{(2)}(0)$ and the linewidth $\Delta \nu$ in fig.~\ref{scaling}acd the pumping rate is $10^{-1}\Gamma_0$ and the interatomic distances are chosen along the white circles in fig.\ 3. The linewidth $\Delta \nu$ is well below $\Gamma_0$ with $N\geq 4$ emitters in the ring and for distances $d$ where $g^{(2)}(0) \approx 1$ but reaches a minimum which is above $\Gamma_0/2$.

\subsection{Computing the Spectrum}
The spectrum we use in order to compute the linewidth via its FWHM is given by the Fourier Transform of the first-order correlation function. Similarly to the second-order correlation function, in the far-field this expression becomes independent of the geometry and is given by
\begin{align}
g^{(1)}(\tau) = \sum_{i,j}\langle\sigma_i^+\sigma_j^-\rangle.
\end{align}

The spectrum can then be computed via~\cite{carmichael2009open} as
\begin{align}\label{eq:spectrum}
S(\omega) = 2 \Re \Bigg\{ \int_0^\infty d\tau e^{-i\omega \tau}\sum_{i,j}\langle \sigma_i^+(\tau) \sigma^-_j \rangle \Bigg\}.
\end{align}
\vfill
\end{document}